\documentstyle[12pt,blois]{article}
\def\mum{\,\mu{\rm m}}
\begin{document}
\heading{
A SUBMILLIMETRE SURVEY OF THE {\em HUBBLE
DEEP FIELD}: UNVEILING DUST-ENSHROUDED
STAR FORMATION IN THE EARLY UNIVERSE}

\author{D.H. Hughes$^{1}$ \& the UK Submillimetre Survey Consortium
\footnote[2]{Consortium membership as in [12]}} 
{$^{1}$ Institute for Astronomy, University of Edinburgh, U.K.}

\begin{bloisabstract}
The advent of sensitive sub-mm array cameras now allows a proper
census of dust-enshrouded massive star-formation
in very distant galaxies, previously hidden activity to which even the 
deepest optical images are insensitive. We present the deepest sub-mm
survey, 
taken with the SCUBA camera on the James Clerk Maxwell
Telescope (JCMT) and centred on the {\em Hubble Deep Field} (HDF). The high 
source density on this 
image implies that the survey is confusion-limited 
below a flux density of 2 mJy. However within the central 80 arcsec radius
independent analyses yield 5 reproducible sources with $S_{850\mu m} >
2$\,mJy which simulations indicate can be ascribed to individual galaxies. 
These data lead to integral 
source counts which are completely inconsistent with a no evolution model, 
whilst the combined brightness of the 5 most 
secure sources in our map is sufficient to account for 30--50\% of the 
previously unresolved sub-mm background, and statistically 
the entire background is resolved at about the 0.3\,mJy level. 
Four of the five brightest sources appear to be associated with galaxies 
which lie in the redshift range $2 \leq z < 4$. With the caveat that
this is a small
sample of sources detected in a small survey area, these submm data 
imply a star-formation density
over this redshift range that is at least five times higher than that 
inferred from the rest-frame ultraviolet output of HDF galaxies.

\end{bloisabstract}

\section{Star-Formation at High Redshift}
 
The
global star-formation history of the Universe \cite{1}, \cite{2}, \cite{3},
derived from optical-UV data, imply that the star-formation and
metal-production rates were $\sim 10$ times greater at $z \simeq 1$
than in the local Universe \cite{4}, that they peaked at $z \simeq 1 - 1.5$ and that they declined to values
comparable to those observed at the present day at $z \simeq 4$.
However these optical-UV data may offer 
a distorted view of the evolution of the
early Universe because the star-formation rate
(SFR) in high-redshift objects is inevitably
under-estimated unless some correction for dust obscuration is
included in deriving the rest-frame UV luminosity \cite{5}.  
Second, it is
possible that an entire population of heavily dust-enshrouded
high-redshift objects, as expected in some models of elliptical galaxy
formation \cite{6}, have gone undetected in the optical/UV surveys.

At high redshifts ($z > 1$), the strongly-peaked far-infrared (FIR)
 radiation emitted by star-formation regions in distant galaxies is
 redshifted into the sub-mm waveband, and the resulting 
 large negative K--correction  at submm wavelengths is
 sufficient  to offset the dimming of galaxies
 due to their cosmological distances.  Consequently the flux density
 of a galaxy at $\lambda \simeq 850\mum$ with fixed intrinsic FIR
 luminosity is expected to be roughly constant at all redshifts in the
 range $1 \leq z \leq 10$ \cite{7}, \cite{8}, \cite{9}.

 Using the new sub-mm array camera SCUBA
 \cite{10} on the 15-m JCMT it is now possible to conduct unbiased sub-mm
 selected surveys and quantify the amount of star-formation activity
 in the young Universe by observing directly the rest-frame FIR
 emission from dust in high-redshift galaxies.  
In this paper we briefly summarise the first
 results from the deepest sub-mm SCUBA survey, complete to a flux density limit
 $S_{850\mum}>2$~mJy, centred on the HDF.  In contrast to mid-IR
 studies \cite{11}, such an 850$\mum$ survey is predicted to be completely
 dominated by sources at $z \ge 1$, and the number of detectable
 sources is very sensitive to the high-redshift evolution of the dusty
 starburst population.  
A full description of these
 data, observing methods, reduction techniques, detailed analysis and
 interpretation is described elsewhere \cite{12}.

\section{A deep sub-mm survey of the {\em Hubble Deep Field}}
The 850$\mu$m SCUBA survey of the HDF, 
with 14.7 arcsecs resolution,  reaches a 1$\sigma$
noise level of 0.45~mJy/beam and represents 
the deepest sub-mm map ever taken.  
The positions of the brightest (and most secure) 5 sources are 
given in table 1, together with their 850-$\mu$m flux densities.
The two major results from this preliminary analysis concern 
the sub-mm background and the star-formation density in the early Universe
\cite{12}. These are discussed in turn.

\subsection{Resolving the submillimetre background}
A P(D) analysis indicates the flux density distribution is
 matched reasonably well by a source density of about
 $7000\,\rm deg^{-2}$ brighter than 1~mJy,
 which corresponds to the observed density of
 brighter sources, extrapolated with a Euclidean
 count slope, with the major caveat that this
 number assumes an unclustered source distribution.
 The counts must continue to flux densities somewhat fainter
 than 1~mJy, but the present data do not have the sensitivity to 
 estimate where the inevitable break from the Euclidean slope occurs.
 This is best constrained by asking at what flux density the
 extrapolated count exceeds the submm background.  By summing the
 flux densities in table 1, a lower limit to the background
 contributed by discrete submm sources of 20~mJy/5.6 arcmin$^2$ is found,
 equivalent to $\nu I_{\nu} = 1.5 \times 10^{-10}\,\rm Wm^{-2}sr^{-1}$, or 
 approximately half the FIRAS background estimate \cite{13}.
 There is, however, evidence in our data, specifically by continuing the
 deconvolution until the residual noise is statistically
 symmetric, or using the cumulative counts to 1~mJy derived above, 
 that the true background
 contributed by discrete sources may be up to a factor of two higher than
 this, essentially identical to the original FIRAS estimate,
 and consistent with more than 50\% of the revised background estimate 
 at $850\mum$ \cite{14} 
 which suggests $\rm \nu I_{\nu} = 5.0 \pm 4 \times 10^{-10} \,\rm W m^{-2} sr^{-1}$. 
 The faint counts must therefore flatten by a
 flux density of about 0.3~mJy, otherwise even this
 background estimate would be exceeded.

 \begin{table}
 \caption[]{\scriptsize Positions and flux densities for the 5 most reliable 
 sub-mm sources in the HDF with $S_{850}>2$~mJy. 
 The photometric redshifts based on the most likely optical 
counterparts \cite{12}.
Star-formation rates calculated from rest-frame UV (2800{\rm \AA})
and FIR (60$\mum$) luminosities are also given. 
An Einstein-de-Sitter cosmology is assumed. 
}
 \begin{center}
 \begin{tabular}{cccccccc}
 Source    & RA (J2000)  & Dec (J2000) & S$_{850\mum}$ &
$\rm z_{est}$ & $\rm log_{10}\, L_{FIR}$ & \multicolumn{2}{c}{SFR ($h^{-2} \rm M_{\odot}  yr^{-1}$)} \\  
  & & & (mJy) & & $(h^{-2} \rm L_{\odot})$ & UV & FIR \\
 HDF850.1  & 12 36 52.32 & +62 12 26.3 & $7.0 \pm 0.4$ 
& 3.4 & 12.15 & 0.7 & 311 \\ 
 HDF850.2  & 12 36 56.68 & +62 12 03.8 & $3.8 \pm 0.4$  
& 3.8 & 11.87 & 0.2 & 161 \\
 HDF850.3  & 12 36 44.75 & +62 13 03.7 & $3.0 \pm 0.4$  
& 3.9 & 11.76 & 0.1 & 127 \\
 HDF850.4  & 12 36 50.37 & +62 13 15.9 & $2.3 \pm 0.4$ 
& 0.9 & 11.83 & 0.3 & 142 \\
 HDF850.5  & 12 36 51.98 & +62 13 19.2 & $2.1 \pm 0.4$ 
& 3.2 & 11.64 & 0.3 &  95 \\
 \end{tabular}
 \end{center}
 \end{table}

\subsection{The redshift distribution of sub-mm sources in the HDF}
A comparison of the typical spectral energy distribution for starburst 
galaxies with the detections (or upper-limits) of the submm HDF sources
at 7, 15, 450 and 850$\mum$, and at 1.4 and 8.5\,GHz (see \cite{12} and 
references therein)
offers a powerful test of whether the submm sources are likely to be
associated with high or low redshift galaxies. When combined with the 
calculation of the chance probability that the submm source is, in turn, 
associated with each of the nearby optical and radio  
sources (for which spectroscopic or photometric redshifts exist), we find that 
four of the five brightest submm sources appear to be associated with 
starburst galaxies in the redshift range $2 \leq z \leq 4$,
whilst HDF850.4 appears to be associated with a starburst galaxy
at $z \leq 1$. 
A superposition of the submm sources on the HST image of the HDF 
and a detailed discussion of their most probable optical HST 
counterparts are presented by Hughes {\it et al.} \cite{12}.

\subsection{The star formation density in the early Universe}
 
 The redshifts of the sub-mm sources, based on  the suggested optical 
 identifications, are consistent with the expectation that the galaxies
 detected in the 850$\mu$m
 survey of the HDF down to a flux limit $S_{850\mum} = 2$~mJy
 should be dominated by objects at redshifts $z \geq 1$.
 Given this, and the flat flux-density--redshift relation between $z = 1$
 and $z = 10$, the dust-enshrouded SFRs (assuming that heating of the 
dust by an obscured AGN is insignificant) of the 5 brightest galaxies
can be robustly estimated despite their imprecise redshifts.
 By summing the FIR SFRs and dividing by the appropriate 
 cosmological volume, a first, conservative 
 estimate of the level of dust-enshrouded star-formation
 rate in the high redshift Universe can be made using observations that
 are  insensitive to the obscuring effects of dust.
 For illustrative purposes, it is assumed that four of
the five sources 
 lie in the redshift interval $2 < z < 4$, 
in which case a lower-limit to the dust-enshrouded star-formation rate 
density is 
 $0.21 \,h \rm M_{\odot} yr^{-1} Mpc^{-3}$ (assuming $q_0 = 0.5$) 
 at $z \simeq 3$. This datum 
 is compared in figure\,1 with the optically-derived 
 star-formation history of the Universe \cite{3}, the dust-corrected 
 star-formation history predicted from the evolution of radio-loud 
 AGN \cite{15} and that inferred from the 
 metal-production rate as determined from the observed column densities 
 and metallicities in QSO absorbers \cite{16}. 
 
This unique submm survey of unprecedented sensitivity has
identified a population of high-redshift dusty starburst
galaxies which contribute a significant fraction of the 
extragalactic submm background. These data also suggest 
that a significant fraction ($>90\%$) of the star-formation activity in the
early Universe may have been missed in previous optical
studies. Four of the five  brightest submillimetre sources in the HDF alone 
provides a density of dust-enshrouded
 star-formation at $z>2$ which is at least a factor of $\simeq 5$
 greater than that deduced from Lyman-limit systems \cite{3}. 
 The extent to which even this is an under-estimate depends on the number
 of sources fainter than $S_{850} = 2$mJy at comparable redshift.

The immediate challenge is therefore 
to determine 
the true redshift distribution of the submm sources, in particular 
to confirm that galaxies detected in the deepest submm
surveys lie at $z > 1$.  This can be achieved by making
millimetre interferometric measurements, with sub-arcsec positional
errors, of the brightest sub-mm sources, 
and subsequently obtaining optical/IR spectroscopic redshifts of
their unambiguously identified optical, IR or radio counterparts.

\begin{figure}[t]
\begin{center}
\setlength{\unitlength}{1mm}
\begin{picture}(200,100)
\includegraphics{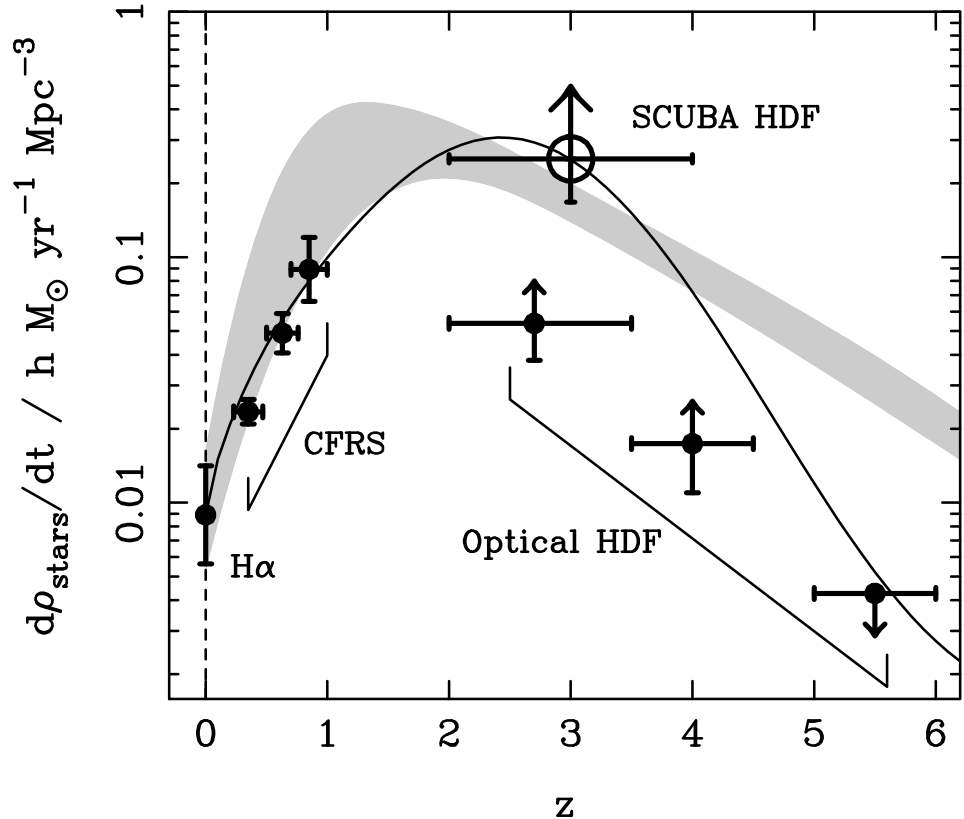}
\end{picture}
\caption[]{\scriptsize 
The global star-formation history of the Universe. Traditionally the
mean {\it co-moving} rate of formation of stars in the Universe,
$d\rho_{\rm stars}/dt$, has been measured from
the total UV luminosity density of galaxies.
At $z<1$, this was measured by the Canada-France
Redshift Survey of Lilly {\it et al.} \cite{1}, and
at higher redshifts from the optical HDF data \cite{3}.
The zero-redshift datum was inferred from local emission-line 
galaxies \cite{4}. The shaded region shows
the prediction (assuming $h=0.65$) due to Pei \& Fall \cite{16}
who argued using the observed column densities in QSO absorbers, plus
the low metallicities in these systems, that the star-formation
rate must have peaked between $z=1$ and $z=2$. 
The solid line illustrates what would
happen if the star-formation rate tracked the total output
of radio-loud AGN \cite{15}. Based on the evidence which
indicates that four of the five brightest sub-mm HDF sources
lie in $2 < z < 4$, we infer a rate about 5 times higher than
that obtained by Madau \cite{3}, but in good agreement with the
external predictions of the rate at these epochs.
}
\end{center}
\end{figure}

 

%
\vfill
\end{document}